\documentclass{ws-ijmpd}

\usepackage{amssymb,amsmath} 
\usepackage{epsfig,array} 

\newcommand{\beqs}{\begin{equation*}}
\newcommand{\beq}{\begin{equation}}

\newcommand{\eeqs}{\end{equation*}}
\newcommand{\eeq}{\end{equation}}

\newcommand{\beqas}{\begin{eqnarray*}}
\newcommand{\beqa}{\begin{eqnarray}}

\newcommand{\eeqas}{\end{eqnarray*}}
\newcommand{\eeqa}{\end{eqnarray}}

\newcommand{\eq}[2]{\begin{equation} #1 \label{#2} \end{equation}}

\newcommand{\al}{\alpha}
\newcommand{\be}{\beta}
\newcommand{\ga}{\gamma}

\newcommand{\si}{\sigma}

\newcommand{\De}{\Delta}

\newcommand{\La}{\Lambda}

\newcommand{\blist}{\begin{itemize}}

\newcommand{\elist}{\end{itemize}}

\providecommand{\href}[2]{#2}

\DeclareFontFamily{OT1}{rsfs}{}
\DeclareFontShape{OT1}{rsfs}{m}{n}{ <-7> rsfs5 <7-10> rsfs7 <10->rsfs10}{} 
\DeclareMathAlphabet{\mycal}{OT1}{rsfs}{m}{n}

\DeclareMathOperator{\extdm}{d}
\newcommand{\extd}{\extdm \!}

\begin{document}

\markboth{D.~GRUMILLER AND F.~PREIS}
{RINDLER FORCE AT LARGE DISTANCES}

\catchline{}{}{}{}{}

\title{Rindler force at large distances}

\author{D.~GRUMILLER AND F.~PREIS}

\address{Institute for Theoretical Physics, Vienna University of Technology\\
Wiedner Hauptstrasse 8-10/136, A-1040 Vienna, Austria\\
E-mail: {\tt grumil@hep.itp.tuwien.ac.at}, {\tt fpreis@hep.itp.tuwien.ac.at}}

\maketitle

\begin{history}
TUW--11--15, {\tt 1107.2373 [astro-ph.co]}

\received{24 May 2011}
\accepted{2 November 2011}
\comby{D.V.~Ahluwalia-Khalilova}
\end{history}

\begin{abstract}
Given some assumptions it is possible to derive the most general post-general relativistic theory of gravity for the distant field of a point mass.
The force law derived from this theory contains a Rindler term in addition to well-known contributions, a Schwarzschild mass and a cosmological constant.
The same force law recently was confronted with solar system precision data.
The Rindler force, if present in Nature, has intriguing consequences for gravity at large distances.
In particular, the Rindler force is capable of explaining about 10\% of the Pioneer anomaly and simultaneously ameliorates the shape of galactic rotation curves.
\end{abstract}

\keywords{IR modifications of gravity; Rindler acceleration; galactic rotation curves; Pioneer anomaly}

\paragraph{}
What is the most general theory of gravity at large distances?
This is an interesting question, whose answer may help to understand some of the puzzles that gravity poses, including the issues of dark matter and dark energy. 

This question was answered recently~\cite{Grumiller:2010bz}.
Of course, any such answer is only as good as the assumptions used in its derivation.
The assumptions of Ref.~\cite{Grumiller:2010bz} were diffeomorphism invariance, spherical symmetry at large distances (which effectively reduces the theory to two dimensions), power counting renormalizability, cosmic censorship at large distances, local validity of Newton's law (based on the tight bounds of Ref.~\cite{Talmadge:1988qz}) and analyticity. 
The first two assumptions imply that gravity at large distances can be described by line-elements of the form
\eq{
\extd s^2=g_{\al\be}\,\extd x^\al\extd x^\be+\Phi^2\,\big(\extd\theta^2+\sin^2\theta\,\extd\phi^2\big)\,.
}{eq:rind1}
The main burden of Ref.~\cite{Grumiller:2010bz} was to construct the most general model that determines the dynamics of the 2-dimensional metric $g_{\al\be}(x^\ga)$ and the surface radius $\Phi(x^\ga)$. 
Exploiting the remaining assumptions it was shown that this model is described by a specific 2-dimensional dilaton gravity action $S$ depending on two constants, $\La$ and $a$ (we use Planck units)
\eq{
S = -\int\!\extd^2x\sqrt{-g}\,\Big[\Phi^2R+2(\partial\Phi)^2-6\Lambda\Phi^2+8a\Phi+2\Big]\,.
}{eq:rind6}
The most general solution to the equations of motion descending from the action \eqref{eq:rind6} was found using the gauge theoretic formulation based upon Ref.~\cite{Cangemi:1992bj}~\footnote{%
A Birkhoff-like theorem can be proven for all 2-dimensional dilaton gravity models, in the following sense.
All classical solutions to the equations of motion exhibit either one or three Killing vectors, see Ref.~\cite{Grumiller:2002nm} and Refs.~therein.
In the present case for non-zero $a$ or $M$ there is exactly one Killing vector.
It is time-like in the relevant causal patch, as evident from the line-element \eqref{eq:rind8}.
}.
In Schwarzschild gauge the surface radius is given by $\Phi = r$, while the 2-dimensional line-element reads
\eq{
g_{\al\be}\,\extd x^\al\extd x^\be = -K^2\extd t^2+\frac{\extd r^2}{K^2} \qquad 
K^2 = 1-\frac{2M}{r}-\Lambda r^2+2ar\,.
}{eq:rind8}
For $M=\La=0$ the line-element \eqref{eq:rind8} is the 2-dimensional Rindler metric. 
The effective field theory \eqref{eq:rind6} therefore differs only in one aspect from spherically symmetric general relativity: 
it permits a Rindler term. 


Solar system data are typically the graveyard of modified theories of gravity~\cite{Olmo:2005zr,Martin:2005bp}.
It is thus pivotal to confront the line-element \eqref{eq:rind1}, \eqref{eq:rind8} with solar system precision data.

To this end we study geodesics on such backgrounds and obtain the standard expressions for angular velocity, $\dot\phi = \ell/r^2$, and radial velocity,
$\dot r^2/2 + V^{\rm eff} = E$,
with some energy $E$ and angular momentum $\ell$.
For time-like test-particles the effective potential derived from the line-element \eqref{eq:rind1}, \eqref{eq:rind8} is given by
\eq{
V^{\rm eff} = -\frac{M}{r} + \frac{\ell^2}{2r^2} - \frac{M\ell^2}{r^3} - \frac{\La r^2}{2} + ar\,\big(1+\frac{\ell^2}{r^2}\big) \,.
}{eq:rind11}
For light-like test-particles only the terms proportional to $\ell^2$ remain.
The first term in $V^{\rm eff}$ is the Newton potential, the second the centrifugal barrier, the third the general relativistic correction, the fourth a cosmic acceleration and the last term proportional to the Rindler acceleration is novel. 
If the  angular momentum vanishes the force $F$ (per unit mass) derived from the effective potential \eqref{eq:rind11} reduces to a Newtonian result, but with Rindler term.
(As a simplification we set the cosmological constant to zero, $\La=0$.) 
\eq{
F = -\frac{M}{r^2} - a
}{eq:P5}

Before we proceed it is important to mention a caveat. 
Namely, in order to trust the effective model at large distances \eqref{eq:rind6} the condition
\eq{
\frac{m}{r_0} \lesssim ar
}{eq:condition}
should hold at least approximately, where $r_0$ is the size of the test-mass. 
Otherwise the self-energy of the test-particle would dominate over the Rindler energy and the picture above fails because the test-particle backreacts appreciably on the ``background''. 
In table \ref{tab:2} some typical systems are considered and it is clarified to which of them \eqref{eq:P5} is applicable.
\begin{table}
\begin{center}
\hspace*{-0.5truecm}\begin{tabular}{|l|l|l|l|l||l|c|}
\hline
Source & Test object & Mass $m$ & Size $r_0$ & Distance $r$ & Ratio $\frac{arr_0}{m}$ & \eqref{eq:P5} ok? \\ \hline
Sun & Pioneer & $10^{11}$ & $10^{35}$ & $10^{47}$ & $10^{9}$ & Yes\\ 
Sun & Earth & $10^{32}$ & $10^{41}$ & $10^{46}$ & $10^{-7}$ & No\\ 
Milky Way & Sun & $10^{38}$ & $10^{44}$ & $10^{55}$ & $10^{-1}$ & Perhaps\\ 
\hline
\end{tabular}
\end{center}
\caption{Checking the inequality \eqref{eq:condition} for various systems with $a$ from \eqref{eq:a}}
\label{tab:2}
\end{table}


In collaboration with ESA we found recently that the best solar system constraint complying with the condition \eqref{eq:condition} comes from radar echo delay~\cite{Carloni:2011ha}. 
More specifically, we calculated the (coordinate) time delay $\De t$ due to light bending and clock effects for a radar signal sent from Earth to some planet or space craft and reflected back to Earth when Earth and the target are in opposition.
To leading order we obtained the following time-delay formula.
\eq{
\De t = 4M \Big(\ln{\frac{4r_E r_T}{r_0^2}}+1\Big) - 2a (r^2_E + r^2_T) 
}{eq:ct11}
Here $r_0$ is a radius of the order of the solar radius, while $r_E$ and $r_T$ are the semi-major axes of Earth and the target, respectively.
The first term on the right hand side of \eqref{eq:ct11} is the general relativistic result, while the second term is the leading correction from the Rindler force.

The Cassini spacecraft data provide a strong bound on corrections to the general relativistic result of time-delay~\cite{Bertotti:2003}.
Exploiting these data we converted the result \eqref{eq:ct11} into a constraint on the magnitude of Rindler acceleration~\cite{Carloni:2011ha}.
\eq{
|a|<5\cdot 10^{-61}\approx 3\cdot 10^{-9}\, m/s^2
}{eq:ct13}
This bound may be improved with future missions like e.g.~the Juno mission by NASA or the EJSM-Laplace mission by ESA.


The model \eqref{eq:rind6} for gravity at large distances predicts the possibility of a Rindler force, but does not determine its sign or magnitude.
We assume now that $a$ is a universal constant and postulate
\eq{
a \approx 10^{-62} \approx 10^{-10}\, m/s^2\,.
}{eq:a}
This value coincides with the critical acceleration in modified Newton dynamics (see for instance Ref.~\cite{Milgrom:2011qt}).
The sign was chosen such that the Rindler term in the Newtonian limit \eqref{eq:P5} produces a force towards the source.
Note that the choice \eqref{eq:a} is compatible with (and not very far from) the experimental bound \eqref{eq:ct13}.
We take \eqref{eq:a} as a working hypothesis and investigate its implications.

Let us discuss observational consequences of the Rindler contribution to the force law \eqref{eq:P5} with the choice above \eqref{eq:a}.
Several of the solar system tests involve the planets as test masses. 
As evident from table \ref{tab:2} the gravitational self-energy is so large that the force law \eqref{eq:P5} is not applicable to them.
In order to find a test mass which would allow to apply \eqref{eq:P5} the ratio $m/r_0$ must not exceed $10^{-17}$ [about 10 Astronomical Units times the quantity $a\approx 10^{-62}$ taken from \eqref{eq:a}]. This is the case for satellites or spacecrafts, which typically have $m/r_0\approx 10^{-24}$, so those objects are prime candidates to test \eqref{eq:P5}. 
As an example we consider the Pioneer spacecrafts. 
The force law \eqref{eq:P5} yields
\eq{
F_{\rm Pioneer} \approx -\frac{10^{38}}{r^2}-10^{-62}\,,
}{eq:P58}
where the force per unit mass is directed towards the Sun for both terms. 
Thus, an anomalous acceleration towards the Sun is predicted. 
Actual experiments found a slightly larger acceleration of about $10^{-61}$ (see Ref.~\cite{Anderson:1998jd}). 
If taken at face value the result \eqref{eq:P58} implies that the Pioneer acceleration observed is too large and should be 90\% artifact and 10\% effect.


\begin{figure}[h!]
\centering
\epsfig{file=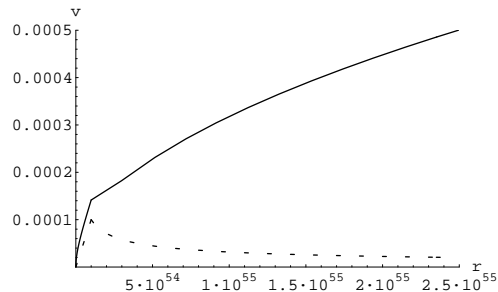,width=0.4\linewidth}
\caption{Rotation curve for dwarf galaxy ($3\,\textrm{kiloparsec}\approx 10^{55}$)}
\label{fig:1}
\end{figure}
\begin{figure}[h!]
\centering
\epsfig{file=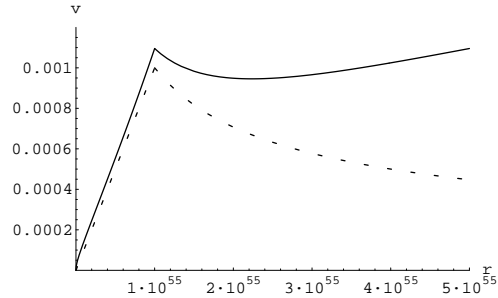,width=0.4\linewidth}
\caption{Rotation curve for large spiral galaxy}
\label{fig:2}
\end{figure}
Consider now galaxies. 
The galactic rotation curve predicted from \eqref{eq:P5} and \eqref{eq:a} is
\eq{
v(r)\approx\sqrt{\frac{M(r)}{r}+10^{-62}\,r}\,.
}{eq:angelinajolie}
For small galaxies ($M\approx 10^8$ solar masses, assuming for simplicity a constant density until $r\approx 10^{54}$ and vanishing density thereafter) we plot the velocity $v$ as a function of the radius $r$ in figure \ref{fig:1}. 
Quantitatively similar curves have been observed for dwarf galaxies. 
The Newtonian prediction without Rindler force corresponds to the dotted line in both figures.
For large galaxies ($M\approx 10^{11}$ solar masses, assuming for simplicity a constant density until $r\approx 10^{55}$ and vanishing density thereafter) the rotation curve is depicted in figure \ref{fig:2}. 
Quantitatively similar curves have been observed for large spiral galaxies. 
Note the maximal velocity of about $10^{-3}$ ($300\,km/s$), close to observational bounds~\cite{Sofue:2000jx}.
It is intriguing that the Newtonian approximation \eqref{eq:P5} of the model \eqref{eq:rind6} appears to be capable of describing gravity at large distances better than the Newtonian limit of spherically symmetric general relativity.


The model \eqref{eq:rind6} shows that a Rindler term leading to the quasi-Newtonian force law \eqref{eq:P5} {\em can} emerge at large distances.
However, it does not --- and cannot --- explain why a Rindler force {\em should} emerge in the first place.
We mention finally scenarios that lead to a Rindler force.

Any modified theory of gravity that at large distances leads to spherically symmetric metrics solving 
$\nabla_\si \nabla^\si R = 0$ or $\nabla_\si \nabla^\si R_{\mu\nu} = 0$
automatically predicts the possibility of a Rindler force.
This includes the theory of conformal Weyl gravity~\cite{Mannheim:1988dj} and, for some critical tuning, the fourth derivative theory introduced recently by L\"u and Pope~\cite{Lu:2011zk}. 
It remains to be seen whether any of these models is capable of passing all theoretical consistency tests and survives all observational bounds.
Any dark matter model that predicts density $\rho=-\frac{a}{2\pi r}$, radial pressure $p_r=-\rho$ and tangential pressure $p_\perp=\frac12\,p_r$ leads to the same dynamics as a Rindler force would.
It remains to be seen whether such exotic dark matter can be derived from some reasonable model.
Finally, it is conceivable that quantum effects within general relativity lead to a Rindler force. 
The first hint in this direction comes from the gravitational scattering of scalar s-waves, where the intermediate geometries where found to contain a Rindler term~\cite{Fischer:2001vz}.
It would be interesting to derive a Rindler force at large distances within more comprehensive approaches to quantum gravity.

\section*{Acknowledgments}

We thank Sante Carloni for collaboration, the head of the ESA Advanced Concepts Team, Leopold Summerer, for encouragement and the following persons for discussions at various stages of the development of the ideas put forward in this essay:
Herbert Balasin, Stanley Deser, Alan Guth, Lorenzo Iorio, Roman Jackiw, Wolfgang Kummer, Philip Mannheim, Stacy McGaugh, Tonguc Rador, Dimitri Vassilevich and Richard Woodard. 

DG was supported by the START project Y435-N16 of the Austrian Science Fund (FWF). 
FP was supported by the FWF project P22114-N16. 
DG and FP acknowledge support from the ESA Advanced Concept Team project Ariadna ID 07/1301, AO/1-5582/07/NL/CB.


\begin{thebibliography}{10}

\bibitem{Grumiller:2010bz}
D.~Grumiller, ``{Model for gravity at large distances},'' {\em Phys. Rev.
  Lett.} {\bf 105} (2010) 211303,
  \href{http://www.arXiv.org/abs/1011.3625}{{\tt 1011.3625}}; Erratum ibid. {\bf 106} (2011) 039901.

\bibitem{Talmadge:1988qz}
C.~Talmadge, J.~P. Berthias, R.~W. Hellings, and E.~M. Standish, ``Model
  independent constraints on possible modifications of {N}ewtonian gravity,''
  {\em Phys. Rev. Lett.} {\bf 61} (1988)
1159--1162.

\bibitem{Cangemi:1992bj}
D.~Cangemi and R.~Jackiw, ``Gauge invariant formulations of lineal gravity,''
  {\em Phys. Rev. Lett.} {\bf 69} (1992) 233--236,
\href{http://arXiv.org/abs/hep-th/9203056}{{\tt hep-th/9203056}}.

\bibitem{Grumiller:2002nm}
D.~Grumiller, W.~Kummer, and D.~V. Vassilevich, ``Dilaton gravity in two
  dimensions,'' {\em Phys. Rept.} {\bf 369} (2002) 327--429,
\href{http://arXiv.org/abs/hep-th/0204253}{{\tt hep-th/0204253}}.

\bibitem{Olmo:2005zr}
G.~J. Olmo, ``{The Gravity Lagrangian according to solar system experiments},''
  {\em Phys. Rev. Lett.} {\bf 95} (2005) 261102,
  \href{http://www.arXiv.org/abs/gr-qc/0505101}{{\tt gr-qc/0505101}}.

\bibitem{Martin:2005bp}
J.~Martin, C.~Schimd, and J.-P. Uzan, ``{Testing for $w<-1$ in the solar
  system},'' {\em Phys. Rev. Lett.} {\bf 96} (2006) 061303,
  \href{http://www.arXiv.org/abs/astro-ph/0510208}{{\tt astro-ph/0510208}}.

\bibitem{Carloni:2011ha}
S.~Carloni, D.~Grumiller, and F.~Preis, {\em Phys. Rev.} {\bf D83} (2011) 124024,
\href{http://www.arXiv.org/abs/1103.0274}{{\tt 1103.0274}}.

\bibitem{Bertotti:2003}
B.~Bertotti, L.~Iess, and P.~Tortora, ``{A test of general relativity using
  radio links with the Cassini spacecraft},'' {\em Nature} {\bf 475} (2003)
  374.

\bibitem{Milgrom:2011qt}
M.~Milgrom, ``{Gravitational Cherenkov losses in MOND theories},'' {\em
  Phys.Rev.Lett.} {\bf 106} (2011) 111101,
  \href{http://www.arXiv.org/abs/1102.1818}{{\tt 1102.1818}}.

\bibitem{Anderson:1998jd}
J.~D. Anderson {\em et al.}, ``{Indication, from Pioneer 10/11, Galileo, and
  Ulysses Data, of an Apparent Anomalous, Weak, Long-Range Accelerattion},''
  {\em Phys. Rev. Lett.} {\bf 81} (1998) 2858--2861,
\href{http://www.arXiv.org/abs/gr-qc/9808081}{{\tt gr-qc/9808081}}.

\bibitem{Sofue:2000jx}
Y.~Sofue and V.~Rubin, ``{Rotation Curves of Spiral Galaxies},'' {\em Ann. Rev.
  Astron. Astrophys.} {\bf 39} (2001) 137--174,
\href{http://www.arXiv.org/abs/astro-ph/0010594}{{\tt astro-ph/0010594}}.

\bibitem{Mannheim:1988dj}
P.~D. Mannheim and D.~Kazanas, ``Exact vacuum solution to conformal weyl
  gravity and galactic rotation curves,'' {\em Astrophys. J.} {\bf 342} (1989)
635--638.

\bibitem{Lu:2011zk}
H.~L{\"u} and C.~Pope, ``{Critical Gravity in Four Dimensions},''
{\em  Phys. Rev. Lett.} {\bf 106} (2011) 181302,
  \href{http://www.arXiv.org/abs/1101.1971}{{\tt 1101.1971}}. 

\bibitem{Fischer:2001vz}
P.~Fischer, D.~Grumiller, W.~Kummer, and D.~V. Vassilevich, ``S-matrix for
  s-wave gravitational scattering,'' {\em Phys. Lett.} {\bf B521} (2001)
  357--363, \href{http://arXiv.org/abs/gr-qc/0105034}{{\tt gr-qc/0105034}}.
Erratum ibid. {\bf B532} (2002) 373.

\end{thebibliography}

\providecommand{\href}[2]{#2}\begingroup\raggedright\endgroup

\end{document}